\documentstyle[prd,preprint,tighten,aps,amssymb,newlfont,%
subfigure,epsfig]{revtex}
\begin{document}
\draft
\preprint{
\begin{tabular}{r}
   UWThPh-1999-65\\
   October 1999
\end{tabular}
}
\title{PHENOMENOLOGY OF\\ NEUTRINO MASSES AND MIXING
\thanks{Talk presented at XXIII International School of
Theoretical Physics on Recent Developments in Theory of Fundamental
Interactions, Ustro\'n, Poland, September 15--22, 1999}}
\author{W. Grimus}
\address{Institute for Theoretical Physics, University of Vienna\\
Boltzmanngasse 5, A--1090 Vienna, Austria}
\maketitle
\begin{abstract}
We discuss all possible schemes
with four massive neutrinos
inspired by the existing experimental indications
in favour of neutrino mixing, namely the atmospheric, solar and LSND
neutrino experiments.
We argue that the scheme with a neutrino
mass hierarchy is not compatible with
the experimental results, likewise
all other schemes with the masses of three
neutrinos close together and the fourth mass separated by a gap needed
to incorporate the LSND neutrino oscillation result.
Only two schemes with two pairs of neutrinos with nearly degenerate masses
separated by this gap of the order of 1 eV
are in agreement with the results
of all experiments, including those where no indications for neutrino
oscillations have been found.
We also point out the possible effect of big-bang nucleosynthesis on
the 4-neutrino mixing matrix and its consequences for neutrino
oscillations. Finally, we study predictions for neutrino oscillation
experiments and $^3$H and $(\beta \beta)_{0\nu}$ decays, following
from the two 
favoured neutrino mass spectra and mixing schemes. These predictions
can be conceived as checks of the input used for arriving at the two
favoured schemes.
\end{abstract}

\section*{The scope of this article}

At present there are three indications in favour of neutrino oscillations: the
results of the solar \cite{Homestake,Kam-sun,GALLEX,SAGE,SK-sun}, 
atmospheric \cite{Kam-atm,IMB,Soudan,SK-atm,SK-mu,SK-stopmu,scholberg,MACRO}
and LSND experiments \cite{LSND-DAR,LSND-DIF,LSND-www}. These three
indications require three different scales of mass-squared differences and,
therefore, four neutrinos with definite mass. Since the LEP
experiments \cite{LEP}
have shown that the number of active neutrinos is three, a
fourth neutrino is needed which is sterile, i.e., its couplings to the $W$
and $Z$ bosons are zero or negligible.

In this paper we take all three indications seriously and thus discuss the
phenomenological analysis of all existing neutrino data in terms of 
\begin{center}
\emph{3 active + 1 sterile neutrino.} 
\end{center}
We will consider the topics of the
nature of the possible neutrino mass spectra, constraints on the neutrino
mixing matrix from the oscillation data and from big-bang nucleosynthesis, and
finally checks and consequences ensuing from the favoured neutrino mass
spectra and the associated mixing matrices.

In the following we will use the abbreviations SBL for short-baseline,
LBL for long-baseline and BBN for big-bang nucleosynthesis.

\section{Introduction}

\subsection{Neutrino mixing}

Neutrino masses and neutrino mixing are natural phenomena in
gauge theories extending the Standard Model 
(see, for example, Ref. \cite{gauge}). However, for the time being,
masses and mixing angles cannot be predicted on theoretical grounds
and they are the central subject of the experimental activity in the field
of neutrino physics.

In the general discussion, we assume that there are $n$
neutrino fields with definite flavours and that 
neutrino mixing is described by a $n \times n$ unitary mixing matrix
$U$ \cite{mixing} such that
\begin{equation}
\nu_{{\alpha}L}
=
\sum_{j=1}^{n}
U_{{\alpha}j}
\,
\nu_{jL}
\qquad
(\alpha=e,\mu,\tau,s_1,\dots,s_{n-3}) \,.
\label{mixing}
\end{equation}
Note that the neutrino fields $\nu_{\alpha L}$
other than the three active neutrino flavour fields 
$\nu_{e L}$, $\nu_{\mu L}$, $\nu_{\tau L}$
must be sterile (for a review see Ref. \cite{sterile})
to comply with the result of the LEP measurements of
the number of neutrino flavours.
The fields $\nu_{jL}$ ($j=1,\dots,n$) are the left-handed components
of neutrino fields with definite masses $m_j$. We assume the ordering
$m_1 \leq m_2 \leq \ldots \leq m_n$ for the neutrino masses. 

The most striking feature of neutrino masses and mixing is the
quantum-mechanical effect of neutrino oscillations 
\cite{BP78-87} (for a review on the early years of neutrino
oscillations see Ref. \cite{samoil}). The transition ($\alpha \neq \beta$) or 
survival ($\alpha = \beta$) probability for
$\nu_\alpha \rightarrow \nu_\beta$ is given by
\begin{equation} \label{prob}
P_{\nu_\alpha\rightarrow\nu_\beta} =
\left|
\sum_{j=1}^n
U_{\beta j}
U_{\alpha j}^*
\exp\!\left(-i\frac{\Delta{m}^{2}_{j1} L}{2E}\right) 
\right|^2 \,,
\end{equation}
where $\Delta{m}^{2}_{jk} \equiv m^2_j - m^2_k$, $L$ is the distance
between neutrino source and detector and $E$ is the neutrino
energy. Eq.(\ref{prob}) is valid for ultrarelativistic neutrinos with
$m^2_j/E^2 \ll 1$ ($j=1,\ldots,n$). There are additional conditions depending
on the neutrino production and detection processes which must hold for
the validity of Eq.(\ref{prob}). See, e.g., Ref.~\cite{osc} and
references therein. 

Let us indicate some important features of Eq.(\ref{prob}):
\begin{itemize}
\item 
The oscillation probability $P_{\bar\nu_\alpha\rightarrow\bar\nu_\beta}$ for
antineutrinos is obtained from $P_{\nu_\alpha\rightarrow\nu_\beta}$ by making
the replacement $U \to U^*$.
\item
The probabilities $P_{\nu_\alpha\rightarrow\nu_\beta}$ and
$P_{\bar\nu_\alpha\rightarrow\bar\nu_\beta}$ depend only on mass-squared
differences, which is explicitly shown by the phase factor 
$\exp (i m^2_1 L/2E)$ multiplying the expression within the absolute value in
the probability (\ref{prob}).
\item
The oscillation probabilities $P_{\nu_\alpha\rightarrow\nu_\beta}$ and
$P_{\bar\nu_\alpha\rightarrow\bar\nu_\beta}$ do not distinguish between the
Dirac or Majorana nature of neutrinos. Note that neutrino fields of
different natures cannot mix.
\item
In the oscillation probabilities, phases of the form
\begin{equation}\label{phase}
\frac{\Delta m^2 L}{2E} \simeq 2.53 \times 
\left( \frac{\Delta m^2}{1\, \mbox{eV}^2} \right)
\left( \frac{E}{1\, \mbox{MeV}} \right)^{-1}
\left( \frac{L}{1\, \mbox{m}} \right) 
\end{equation}
occur, where $\Delta m^2$ is a generic mass-squared difference. 
Given $E$ and $L$, these phases
determine the order of magnitude of $\Delta m^2$ a neutrino
oscillation experiment is sensitive to.
\end{itemize}

Let us discuss two examples illustrating the last point.
Clearly, experiments can only see phases (\ref{phase}) if they are not
too small, say if they are of order 1. The first example concerns
SBL reactor experiments. By convention, SBL experiments 
are sensitive to mass-squared differences $\Delta m^2 \gtrsim 0.1$ eV$^2$.
With $E \sim 1$ MeV it follows that $L \gtrsim 10$ m is a sufficient
distance between neutrino source and detector to achieve this
sensitivity. On the other hand, the longest baseline possible on earth
is 13000 km, the diameter of the earth. In this case, the atmospheric
neutrino flux is available with the largest flux around $E \sim 1$
GeV. The requirement that the phase (\ref{phase}) is larger than 0.1
leads to a sensitivity estimate of $\Delta m^2 \gtrsim 10^{-4}$ eV$^2$.

\subsection{Indications in favour of neutrino oscillations}

At present, indications that neutrinos are
massive and mixed have been found in solar neutrino experiments
(Homestake \cite{Homestake},
Kamiokande \cite{Kam-sun},
GAL\-LEX \cite{GALLEX}, SAGE \cite{SAGE} and
Super-Kamiokande \cite{SK-sun}),
in atmospheric neutrino experiments
(Kamiokande \cite{Kam-atm}, IMB \cite{IMB}, Soudan \cite{Soudan},
Super-Kamiokande \cite{SK-atm} and MACRO \cite{MACRO}) and
in the LSND experiment \cite{LSND-DAR,LSND-DIF} 
(see also the review \cite{BGG-review}).
From the analyses of the data of these experiments
in terms of neutrino oscillations
one infers the following scales 
of neutrino mass-squared differences:
\begin{itemize}
\item \textbf{Solar neutrino deficit:} Interpreted as effect of
neutrino oscillations, the relevant value of the mass-squared difference is
determined as \cite{BKS98,GG99}
\begin{equation}
\Delta{m}^2_{\mathrm{solar}}
\sim
10^{-5} \, \mathrm{eV}^2
\, (\mbox{MSW})
\quad \mbox{or} \quad
\Delta{m}^2_{\mathrm{solar}}
\sim
10^{-10} \, \mathrm{eV}^2
\, (\mbox{vac. osc.})
\label{DMsun}
\end{equation}
The two possibilities for $\Delta{m}^2_{\mathrm{solar}}$
correspond, respectively, to the
MSW \cite{MSW} and to the vacuum oscillation
solutions of the solar neutrino problem. The solar neutrino
experiments are $\nu_e$ disappearance experiments.
\item \textbf{Atmospheric neutrino anomaly:} Interpreted as effect of
neutrino oscillations, the zenith angle dependence of the atmospheric
neutrino anomaly \cite{Kam-atm,SK-atm} using the so-called
contained and partially contained multi-GeV events \cite{scholberg} gives
\begin{equation}
\Delta{m}^2_{\mathrm{atm}}
= 3.5 \times 10^{-3} \, \mathrm{eV}^2
\label{DMatm}
\end{equation}
with $\sin^2 2\theta_\mathrm{atm} = 1$ for the mixing angle
as best fit values under the assumption of 
$\stackrel{\scriptscriptstyle (-)}{\nu}_{\hskip-3pt \mu} \to
 \stackrel{\scriptscriptstyle (-)}{\nu}_{\hskip-3pt \tau}$ 
oscillations. 
In essence, the atmospheric neutrino anomaly is interpreted as 
$\stackrel{\scriptscriptstyle (-)}{\nu}_{\hskip-3pt \mu}$ disappearance.
\item \textbf{LSND experiment:} The evidence for 
$\stackrel{\scriptscriptstyle (-)}{\nu}_{\hskip-3pt \mu} \to
 \stackrel{\scriptscriptstyle (-)}{\nu}_{\hskip-3pt e}$ 
oscillations in this experiment leads to \cite{LSND-DAR}
\begin{equation}
\Delta{m}^2_\mathrm{LSND} \sim 1 \, \mathrm{eV}^2 \,.
\label{DMlsnd}
\end{equation}
The result of the LSND experiment is the only evidence for neutrino 
appearance.
\end{itemize}

Thus, due to the three different scales of $\Delta m^2$,
at least four light neutrinos with definite masses
must exist in nature
in order to accommodate the results of all neutrino oscillation
experiments, and because of the LEP result on the number of active
neutrinos the existence of at least one non-interacting sterile 
neutrino is required. In the following, apart from the SBL discussion
in Section \ref{SBL}, we will confine ourselves to four neutrinos. For
early works on four neutrinos see Ref. \cite{four}, for general 
phenomenological discussions see Refs. \cite{BGG96,OY97,BPWW98,BGGS99}.

\section{Types of 4-neutrino mass spectra}

\begin{figure}[t]
\begin{center}
\setlength{\unitlength}{1.0cm}
\begin{tabular*}{0.99\linewidth}{@{\extracolsep{\fill}}cccccc}
\begin{picture}(1,4) 
\thicklines
\put(0.1,0.2){\vector(0,1){3.8}}
\put(0.0,0.2){\line(1,0){0.2}}
\put(0.4,0.15){\makebox(0,0)[l]{$m_1$}}
\put(0.0,0.4){\line(1,0){0.2}}
\put(0.4,0.45){\makebox(0,0)[l]{$m_2$}}
\put(0.0,0.8){\line(1,0){0.2}}
\put(0.4,0.8){\makebox(0,0)[l]{$m_3$}}
\put(0.0,3.5){\line(1,0){0.2}}
\put(0.4,3.5){\makebox(0,0)[l]{$m_4$}}
\end{picture}
&
\begin{picture}(1,4) 
\thicklines
\put(0.1,0.2){\vector(0,1){3.8}}
\put(0.0,0.2){\line(1,0){0.2}}
\put(0.4,0.2){\makebox(0,0)[l]{$m_1$}}
\put(0.0,0.6){\line(1,0){0.2}}
\put(0.4,0.55){\makebox(0,0)[l]{$m_2$}}
\put(0.0,0.8){\line(1,0){0.2}}
\put(0.4,0.85){\makebox(0,0)[l]{$m_3$}}
\put(0.0,3.5){\line(1,0){0.2}}
\put(0.4,3.5){\makebox(0,0)[l]{$m_4$}}
\end{picture}
&
\begin{picture}(1,4) 
\thicklines
\put(0.1,0.2){\vector(0,1){3.8}}
\put(0.0,0.2){\line(1,0){0.2}}
\put(0.4,0.2){\makebox(0,0)[l]{$m_1$}}
\put(0.0,2.9){\line(1,0){0.2}}
\put(0.4,2.9){\makebox(0,0)[l]{$m_2$}}
\put(0.0,3.3){\line(1,0){0.2}}
\put(0.4,3.25){\makebox(0,0)[l]{$m_3$}}
\put(0.0,3.5){\line(1,0){0.2}}
\put(0.4,3.55){\makebox(0,0)[l]{$m_4$}}
\end{picture}
&
\begin{picture}(1,4) 
\thicklines
\put(0.1,0.2){\vector(0,1){3.8}}
\put(0.0,0.2){\line(1,0){0.2}}
\put(0.4,0.2){\makebox(0,0)[l]{$m_1$}}
\put(0.0,2.9){\line(1,0){0.2}}
\put(0.4,2.85){\makebox(0,0)[l]{$m_2$}}
\put(0.0,3.1){\line(1,0){0.2}}
\put(0.4,3.15){\makebox(0,0)[l]{$m_3$}}
\put(0.0,3.5){\line(1,0){0.2}}
\put(0.4,3.5){\makebox(0,0)[l]{$m_4$}}
\end{picture}
&
\begin{picture}(1,4) 
\thicklines
\put(0.1,0.2){\vector(0,1){3.8}}
\put(0.0,0.2){\line(1,0){0.2}}
\put(0.4,0.2){\makebox(0,0)[l]{$m_1$}}
\put(0.0,0.6){\line(1,0){0.2}}
\put(0.4,0.6){\makebox(0,0)[l]{$m_2$}}
\put(0.0,3.3){\line(1,0){0.2}}
\put(0.4,3.25){\makebox(0,0)[l]{$m_3$}}
\put(0.0,3.5){\line(1,0){0.2}}
\put(0.4,3.55){\makebox(0,0)[l]{$m_4$}}
\end{picture}
&
\begin{picture}(1,4) 
\thicklines
\put(0.1,0.2){\vector(0,1){3.8}}
\put(0.0,0.2){\line(1,0){0.2}}
\put(0.4,0.15){\makebox(0,0)[l]{$m_1$}}
\put(0.0,0.4){\line(1,0){0.2}}
\put(0.4,0.45){\makebox(0,0)[l]{$m_2$}}
\put(0.0,3.1){\line(1,0){0.2}}
\put(0.4,3.1){\makebox(0,0)[l]{$m_3$}}
\put(0.0,3.5){\line(1,0){0.2}}
\put(0.4,3.5){\makebox(0,0)[l]{$m_4$}}
\end{picture}
\\
(I) & (II) & (III) & (IV) & (A) & (B)
\end{tabular*}
\end{center}
\caption{ \label{4spectra}
The six types of neutrino mass spectra that can accommodate 
the solar, atmospheric and LSND scales of $\Delta{m}^2$. The different
distances between the masses on the vertical axes symbolize the
different scales of $\Delta{m}^2$. The spectra (I) -- (IV) define
class 1, whereas class 2 comprises (A) and (B).}
\end{figure}
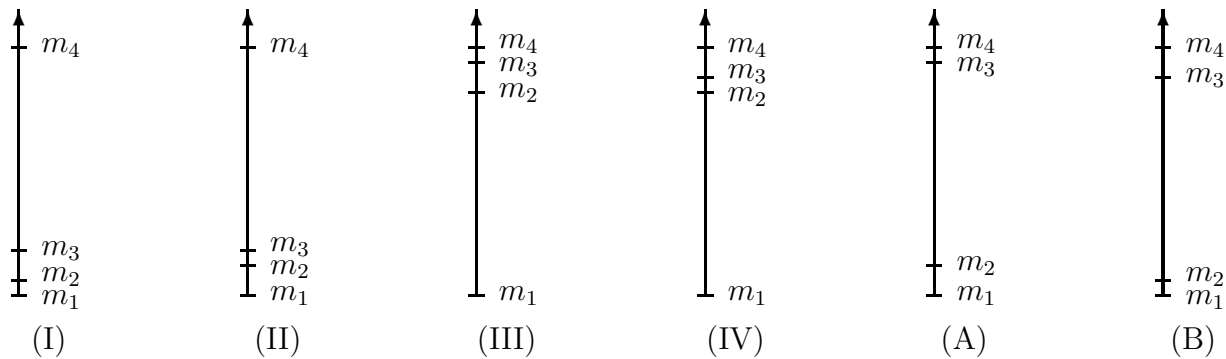

With four massive neutrinos and the ordering
$m_1 < m_2 < m_3 < m_4$ among the masses, there are six possible types
of neutrino mass spectra 
which accommodate the three mass-squared differences required by the
experimental data. In four of them three masses form a cluster
separated by the gap from the fourth mass needed to describe the LSND
experiment (types (I) -- (IV)). Spectrum (I) is the hierarchical type,
Spectrum (III) is sometimes called inverted hierarchy (see
Fig.~\ref{4spectra}). The remaining two spectra denoted by (A) and (B)
have two nearly degenerate mass pairs separated by the LSND gap (see
Fig.~\ref{4spectra}). One of the main focuses of this article is the
discussion of these 6 types of neutrino mass spectra in the light of
all available experimental data.

\section{SBL experiments}
\label{SBL}

The material discussed in this section is independent of the number of
neutrinos. Therefore, this number will be kept general and denoted by $n$.

\subsection{Basic assumption and formalism}
\label{basic}

We will make the following basic assumption \cite{BGG96,pune}
in the further discussion in this report: 
\begin{center}
\emph{A single $\Delta m^2$ is relevant in SBL neutrino experiments.}
\end{center}
This assumption is trivially fulfilled for $n=4$.
In accordance with Eq.(\ref{DMlsnd}) we denote this $\Delta m^2$ by
\begin{equation}\label{DMsbl}
\Delta m^2_\mathrm{LSND} \equiv \Delta m^2_\mathrm{SBL} \,.
\end{equation}

As a consequence of this assumption
the neutrino mass
spectrum consists of two groups of
close masses,
separated by a mass difference in the eV range.
Denoting the neutrinos of the two groups by
$ \nu_1, \ldots , \nu_r $
and
$ \nu_{r+1}, \ldots , \nu_n $,
the mass spectrum looks like
\begin{equation}
m^2_1 \leq \ldots  \leq m^2_r \ll
m^2_{r+1} \leq \ldots \leq m^2_n
\end{equation}
such that
\begin{equation}\label{sbl}
\begin{array}{lcccc}
\Delta m^2_{kj} \ll \Delta m^2_{\mbox{\scriptsize SBL}} 
& \mbox{ for } & 1 \leq j < k \leq r & 
\mbox{ and } & r+1 \leq j < k \leq n, \\[2mm]
\Delta m^2_{kj} \simeq \Delta m^2_{\mbox{\scriptsize SBL}} & 
\mbox{ for } & 1 \leq j \leq r & 
\mbox{ and } & r+1 \leq k \leq n
\end{array}
\end{equation}
holds for the purpose of the SBL formalism. 

Eq.(\ref{prob}) together with Eq.(\ref{sbl}) gives the SBL transition
probability 
\begin{equation} \label{sblprob}
P^{(\mbox{\scriptsize SBL})}_{\nu_{\alpha}\rightarrow\nu_\beta} =
\left|
\sum_{j=1}^r
U_{\beta j}
U_{\alpha j}^* +
\exp\!\left(-i\frac{\Delta m^2_{\mbox{\scriptsize SBL}} L}{2E}\right) 
\sum_{j=r+1}^n
U_{\beta j}
U_{\alpha j}^* \right|^2 \,.
\end{equation}

\subsection{SBL formulas}

For the probability of the transition
$\nu_{\alpha}\rightarrow\nu_{\beta}$
($\alpha\neq\beta$) we obtain from Eq.(\ref{sblprob}) 
\begin{equation} \label{Pa b}
P^{(\mbox{\scriptsize SBL})}_{\nu_{\alpha}\rightarrow\nu_{\beta}}
=
\frac{1}{2}
A_{\alpha;\beta}
\left( 1 - \cos \frac{\Delta m^2_{\mbox{\scriptsize SBL}} L}{2E} \right)
\,,
\end{equation}
where the oscillation amplitude
$A_{\alpha;\beta}$
is given by
\begin{equation} \label{Aab}
A_{\alpha;\beta}  =
4 \left| \sum_{j \geq r+1} U_{\beta j} U_{\alpha j}^* \right|^2
=
4 \left| \sum_{j \leq r}
U_{\beta j} U_{\alpha j}^* \right|^2
\,.
\end{equation}
Eqs.(\ref{Pa b}) and (\ref{Aab}) follow from the unitarity of $U$.
Furthermore, the oscillation amplitude $A_{\alpha;\beta}$ fulfills
the condition $A_{\alpha;\beta} = A_{\beta;\alpha} \leq 1$.
The second part of this equation is a consequence of 
the Cauchy--Schwarz inequality and the unitarity of the mixing matrix.
The survival probability of $\nu_{\alpha}$ is
calculated as
\begin{equation} \label{Pa}
P^{(\mbox{\scriptsize SBL})}_{\nu_{\alpha}\rightarrow\nu_{\alpha}}
=
1 - \sum_{\beta\neq\alpha}
P_{\nu_{\alpha}\rightarrow\nu_\beta}^\mathrm{(SBL)}
=
1 - \frac{1}{2}
B_{\alpha}
\left(1 - \cos \frac{\Delta{m}^2_{\mbox{\scriptsize SBL}} L}{2E} \right)
\end{equation}
with the survival amplitude
\begin{eqnarray} \label{Ba}
B_{\alpha} & = &
4 \left( \sum_{j \geq r+1} |U_{\alpha j}|^2 \right)
\left( 1 - \sum_{j \geq r+1} |U_{\alpha j}|^2 \right) \nonumber \\
& = & 4 \left( \sum_{j \leq r} |U_{\alpha j}|^2 \right)
\left( 1 - \sum_{j \leq r} |U_{\alpha j}|^2 \right) \,.
\end{eqnarray}
Conservation of probability gives the important relation
\begin{equation}\label{BSA}
B_{\alpha} =
\sum_{\beta \neq \alpha}
A_{\alpha;\beta} \leq 1 \,.
\end{equation}

The expressions (\ref{Pa b}) and (\ref{Pa})
describe the transitions between all possible
neutrino states,
whether active or sterile.
Let us stress that with the basic assumption in the beginning of this
subsection the oscillations in
all channels are characterized
by the same oscillation length
\begin{equation}
l_{\mbox{\scriptsize osc}} = 
4 \pi \frac{E}{\Delta m^2_\mathrm{SBL}} =
2.48 \: \mbox{m} \: \left( \frac{E}{1 \: \mbox{MeV}} \right)
\left( \frac{1 \: \mbox{eV}^2}{\Delta m^2_\mathrm{SBL}} \right)
\,. 
\end{equation}
Furthermore, the
substitution $U \to U^*$ in the amplitudes (\ref{Aab}) and (\ref{Ba})
does not change them and therefore it ensues from the basic SBL assumption
that the probabilities (\ref{Pa b}) and (\ref{Pa}) hold for
antineutrinos as well and hence there is no CP violation in SBL
neutrino oscillations.

\subsection{The relation between SBL $n$-neutrino oscillations and\\
2-neutrino oscillations}

The oscillation probabilities (\ref{Pa b}) and (\ref{Pa}) look like
2-flavour probabilities. Defining  
$\sin^2 2\theta_{\alpha\beta} \equiv A_{\alpha;\beta}$,
$\sin^2 2\theta_{\alpha} \equiv B_{\alpha}$ and
$\sin^2 2\theta_{\beta} \equiv B_\beta$ for $\alpha \neq \beta$,
the resemblance is even more striking. It means that the basic SBL
assumption allows to use the 2-flavour oscillation formulas in SBL
experiments.
However, genuine 2-flavour 
$\nu_\alpha \leftrightarrow \nu_\beta$ neutrino oscillations
are characterized by a single mixing angle 
$\theta_{\alpha\beta} = \theta_\alpha = \theta_\beta$.

\begin{figure}[t!]
\begin{center}
\mbox{\epsfig{file=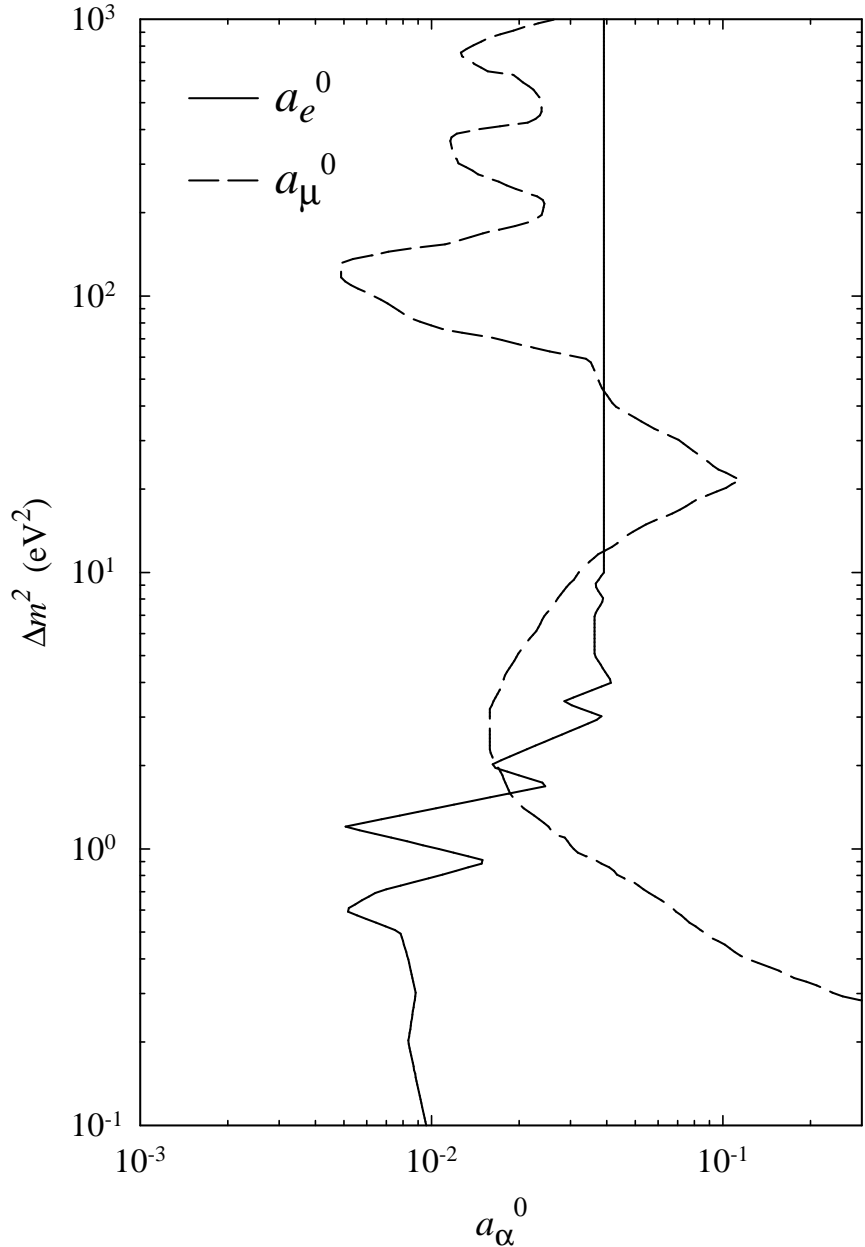,width=0.7\linewidth} \hphantom{xxxxx}}
\end{center}
\caption{\label{a0plot} The bounds $a^0_\alpha$ ($\alpha = e, \mu$).}
\end{figure}

\section{SBL disappearance experiments}

For the two flavours $\alpha = e$ and $\mu$, results of disappearance
experiments are available. We will use the 90\% exclusion plots of the
Bugey reactor experiment \cite{Bugey95} for 
$\bar\nu_e$ disappearance
and the 90\% exclusion plots of the CDHS \cite{CDHS84} and CCFR
\cite{CCFR84} accelerator experiments for $\nu_\mu$
disappearance. Since no neutrino disappearance has been seen in SBL
experiments, there are
upper bounds $B_{\alpha}^0$ on the disappearance amplitudes for
$\alpha = e, \mu$.
These experimental bounds are functions of
$\Delta m^2_{\mbox{\scriptsize SBL}}$. It follows that
\begin{equation}\label{ca}
B_{\alpha} = 4\, c_\alpha (1-c_\alpha) \leq B_{\alpha}^0
\quad \mbox{with} \quad
c_\alpha \equiv \sum_{j=1}^r |U_{\alpha j}|^2
\end{equation}
and, therefore \cite{BBGK},
\begin{equation}
c_\alpha \leq a^0_\alpha \quad \mbox{or} \quad c_\alpha \geq
1-a^0_\alpha \quad \mbox{with} \quad \label{a0}
a^{0}_{\alpha} \equiv \frac{1}{2}
\left(1-\sqrt{1-B_{\alpha}^{0}}\,\right)
\,.
\end{equation}
This equation formulates the important constraints on the mixing
matrix $U$ stemming from the negative results of SBL disappearance
neutrino oscillation experiments. Note that
Eq.(\ref{a0}) shows that $a^0_\alpha \leq 1/2$. Furthermore, since the
upper bounds $B_{\alpha}^0$ on the survival amplitudes $B_\alpha$ are
functions of $\Delta m^2_\mathrm{SBL}$, the same is true for the
bounds $a^0_\alpha$. In the region of $\Delta m^2_\mathrm{SBL}$ where
no experimental restrictions on the survival amplitude 
are available, we have $B_{\alpha}^0 \uparrow 1$, in which case it
follows that $a^0_\alpha \uparrow 0.5$.

In Fig.~\ref{a0plot} the bounds
$a^0_e$ and $a^0_\mu$ are plotted as functions of
$\Delta m^2_\mathrm{SBL}$ in the wide range
$10^{-1}\; \mbox{eV}^2 \leq \Delta m^2_{\mbox{\scriptsize SBL}} \leq 
10^3 \; \mbox{eV}^2$.
In this range $a^0_e$ is small ($a^0_e \lesssim 4 \times 10^{-2}$) and
$a^0_\mu \lesssim 10^{-1}$ for
$\Delta m^2_{\mbox{\scriptsize SBL}} \gtrsim 0.5$ eV$^2$. 

\section{The LSND experiment}

The LSND experiment investigates $\bar\nu_\mu\to\bar\nu_e$
oscillations, where the $\bar\nu_\mu$ flux is generated by $\mu^+$
decay at rest \cite{LSND-DAR}, and $\nu_\mu\to\nu_e$ oscillations, 
where the neutrino source is given by $\pi^+$ decay in flight
\cite{LSND-DIF}. Both channels have shown evidence
in favour of neutrino oscillations with perfectly compatible
results and, therefore, give a non-zero measurement of the transition 
amplitude $A_{\mu;e}$ (\ref{Aab}). On the other hand, from the negative
result of the Bugey experiment and from inequality (\ref{BSA}) we also
have the constraint
\begin{equation}
A_{\mu;e} \leq B^0_e \,.
\end{equation}
From the 90\% CL plot \cite{LSND-www} of the LSND collaboration and
from the Bugey bound $B^0_e$ one obtains approximately
\begin{equation}
2 \times 10^{-3} \lesssim A_{\mu;e} \lesssim 4 \times 10^{-2}
\end{equation}
and
\begin{equation}\label{range}
0.2 \: \mbox{eV}^2 \lesssim \Delta m^2_\mathrm{SBL} \lesssim 
2 \: \mbox{eV}^2 \,.
\end{equation}
See, e.g., also Fig.~5.9 in Ref.~\cite{BGG-review}.
These ranges are also compatible with the negative result of the
KARMEN experiment \cite{KARMEN}.

\section{The Super-Kamiokande up-down asymmetry}

In the Super-Kamiokande experiment -- like before in the Kamiokande
experiment -- atmospheric electron and muon neutrinos are measured by
the Cherenkov light of electrons and muons, respectively, produced by
charged current interactions of the neutrinos. Thus, $e$-like events
appear as diffuse rings and $\mu$-like events as sharp rings in the
detector. A distinguished class of events is given by the single-ring
(1r) events which are fully contained (FC) in the inner
detector. These events are charged current $e$-like and $\mu$-like
events with very high probability \cite{SK-atm}. Partially contained
(PC) events have tracks exiting the inner detector and are nearly
100\% $\mu$-like events. The zenith angle distributions of Kamiokande
\cite{Kam-atm} and Super-Kamiokande \cite{SK-atm}, which gave the
first evidence for atmospheric neutrino oscillations, 
are based on such events, and the up-down asymmetry of 
Super-Kamiokande as well \cite{SK-atm}. 

Note that in the Super-Kamiokande experiment muons going through
\cite{SK-mu} or stopping \cite{SK-stopmu} in the detector are also
measured, which originate from atmospheric muon neutrinos interacting
with the rock beneath the detector. Evaluating these events under the
hypothesis of neutrino oscillations gives results for 
$\Delta m^2_\mathrm{atm}$ and the atmospheric mixing angle compatible
with the oscillation parameters derived from the zenith angle 
distribution \cite{SK-mu,SK-stopmu,scholberg}. The same applies to
the result of the MACRO experiment on through-going muon events
\cite{MACRO}. Moreover, these types
of events, which correspond to neutrino energies of 
$E \sim 100$ GeV for the through-going
and $E \sim 10$ GeV for the stopping events, have the capacity to
allow for a distinction between the $\nu_\mu\to\nu_\tau$ and
$\nu_\mu\to\nu_s$ solutions of the atmospheric neutrino anomaly. At
present, the sterile neutrino solution is disfavoured at about 95\% CL
\cite{nakahata}. 

The zenith angle $\theta_z$ of an $e$-like or $\mu$-like event is 
defined as the angle between the vertical line and the direction of 
the electron or muon track. For multi-GeV events, defined by a visible
energy larger than 1.33 GeV, the average angle between the charged
lepton direction and the neutrino direction is around $20^\circ$
\cite{SK-atm}. Since $\cos \theta_z = \pm 0.2$ corresponds to
$90^\circ \pm 11.5^\circ$ it is reasonable to define up (U) and down
(D) going $\mu$-like events in the following way:
\begin{equation}\label{UD}
\setlength{\arraycolsep}{1mm}
\begin{array}{rl}
U = \# ( \mbox{multi GeV FC 1r + PC $\mu$-like
             events with}& -1 \leq \cos \theta_z \leq -0.2 )\,, \\
D = \# ( \mbox{multi GeV FC 1r + PC $\mu$-like
             events with}&
\hphantom{-}1 \geq \cos \theta_z \geq \hphantom{-}0.2 ) \,.
\end{array}
\end{equation}
Clearly, if there are no neutrino oscillations, we would have $U=D$.
Super-Kamiokande has measured the up-down asymmetry with the latest result
\cite{scholberg} 
\begin{equation}\label{Amu}
A_\mu = \frac{U-D}{U+D} = -0.311 \pm 0.043 \pm 0.01 \,.
\end{equation}
This value constitutes the most compelling evidence for neutrino
oscillations at present.
The error $\pm 0.01$ stems from an estimation of 
the up-down asymmetric effects of the
magnetic field of the earth on the primary cosmic ray flux.
The value for the corresponding asymmetry for $e$-like events (defined
via (\ref{UD}) but without PC) is given by
$A_e = -0.036 \pm 0.067 \pm 0.02$ \cite{SK-atm} and is compatible with
zero, i.e., no oscillations of atmospheric electron neutrinos.

\section{The 4-neutrino mass hierarchy is disfavoured by the data}

In the case of
a neutrino mass hierarchy,
$ m_1 \ll m_2 \ll m_3 \ll m_4 $,
the mass-squared differences $\Delta{m}^{2}_{21}$ and 
$\Delta{m}^{2}_{32} \simeq \Delta{m}^{2}_{31}$ are relevant
for the suppression of the flux of solar neutrinos and for the
atmospheric neutrino anomaly, respectively.
In this case the quantity $c_\alpha$ is defined via $r=3$ 
(see the formalism in Section \ref{basic} and the definition (\ref{ca}))
and, therefore, we have 
\begin{equation}
c_\alpha = \sum_{j=1}^3 |U_{\alpha j}|^2 \,.
\end{equation}
In the following, according to the 4-neutrino assumption, 
we assume that $\Delta m^2_\mathrm{SBL}$ is in the numerical range 
(\ref{range}) given by the result of the LSND experiment.
Our aim is to derive three bounds on $c_\mu$ as functions of 
$\Delta m^2_\mathrm{SBL}$, using as input various oscillation data. We
will finally see that these bound are incompatible with each other,
thus strongly disfavouring the hierarchical neutrino mass spectrum.

The first bound we need is given by Eq.(\ref{a0}):
\begin{equation}\label{amu}
\mbox{Bound a:} \quad
c_\mu \leq a^0_\mu \quad \mbox{or} \quad c_\mu \geq 1-a^0_\mu \,.
\end{equation}
For this bound the experimental input is the data on SBL 
$\stackrel{\scriptscriptstyle (-)}{\nu}_{\hskip-3pt \mu}$ disappearance
\cite{CDHS84,CCFR84}. 

For the derivation of the next bound we refer the reader to
Ref. \cite{BGGS99}. It is based on the up-down asymmetry
\cite{SK-atm,scholberg}: 
\begin{equation}\label{ineq}
\mbox{Bound b:} \quad
-A_\mu \leq \frac{c_\mu^2 + 2\, a^0_e (1-a^0_e) /r}{c_\mu^2 + 2(1-c_\mu)^2} 
\,,
\end{equation}
where $r \equiv n_\mu/n_e$ is defined as the ratio of $\mu$-like to
$e$-like events in the detector without neutrino oscillations. Its
numerical value $r \simeq 2.8$ can be read off from Fig. 3 in
Ref.~\cite{SK-atm}. Because of the smallness of $a^0_e$ the bound
(\ref{ineq}) has a very weak dependence on the precise value of $r$.
For the bound (\ref{ineq}), in addition to $A_\mu$, also SBL
disappearance data \cite{Bugey95} have been used and the lower bound
on the survival probability of solar neutrinos given by \cite{BGKP}
$P^\odot_{\nu_e\to\nu_e} \geq |U_{e4}|^4 = (1-c_e)^2$. The latter
inequality shows that only the possibility $c_e \geq 1-a^0_e$ is
allowed. 

The third bound uses the fact that the LSND result establishes a lower
bound $A_{\mu;e}^\mathrm{min}$ on the transition amplitude $A_{\mu;e}$:
\begin{equation}\label{LSND-bound}
\mbox{Bound c:} \quad
c_\mu \leq 1 - A^\mathrm{min}_{\mu;e}/4a^0_e \,.
\end{equation}
It derives from $A_{\mu;e} = 4 (1-c_e)(1-c_\mu)$ (see
Eq.(\ref{Aab})) and $c_e \geq 1-a^0_e$.

\begin{figure}[t]
\begin{center}
\mbox{\epsfig{file=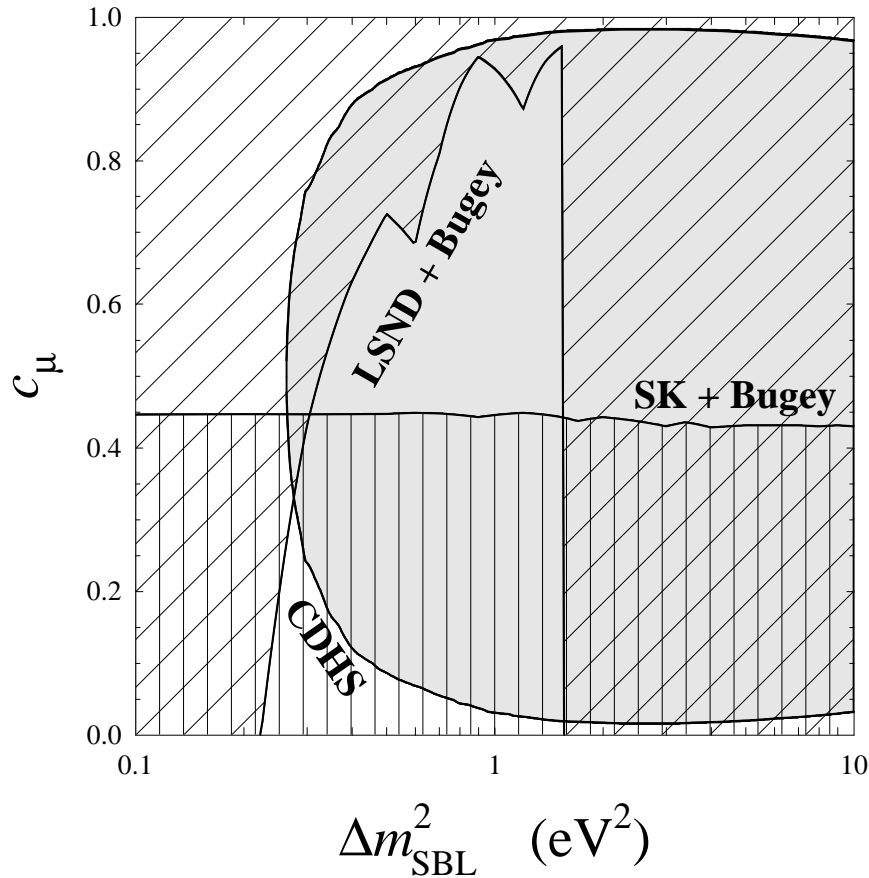,width=0.7\linewidth}}
\end{center}
\caption{ \label{cmu}
Regions in the $\Delta{m}^2_\mathrm{SBL}$--$c_\mu$
plane disfavoured by the results of the CDHS, CCFR,
LSND, Super-Kamiokande and Bugey experiments in the case of the
spectra of class 1. The shaded region is excluded by the inequalities
(\ref{amu}) and the region with oblique hatching
by the bound (\ref{LSND-bound}). The nearly
horizontal curve labelled SK + Bugey represents the lower bound
(\ref{ineq}) derived from the Super-Kamiokande up--down asymmetry. 
No allowed region is left in this plot and, therefore,
the spectra of class 1 are disfavoured by the data.}
\end{figure}

In Fig.~\ref{cmu} the bounds a, b, c, labelled by CDHS, SK+Bugey, 
LSND+\-Bu\-gey, respectively, are plotted in the
$\Delta{m}^2_\mathrm{SBL}$--$c_\mu$ plane. Note that bound c is
practically a horizontal line due to the smallness of the term
containing $a^0_e$. The three bounds, which are
all derived from 90\% CL data, leave no allowed region in the
plot.
Thus the hierarchical mass spectrum (I) is strongly disfavoured
by the data. The same arguments presented here can be used also for
the other spectra (II), (III), (IV) of class 1 (see
Fig.~\ref{4spectra}) by defining $c_\alpha$ (\ref{ca})
through a summation over the indices of the three close masses 
for each of the spectra of class 1 \cite{BGGS99}.

\section{The favoured 4-neutrino mass spectra (A) and (B)}

Now we are left with only 
two possible neutrino mass spectra
in which the four neutrino
masses appear in two pairs
separated by
$ \sim 1 \, \mathrm{eV} $ \cite{BGG96,OY97}:
\begin{equation} \label{AB}
\mbox{(A)}
\quad
\underbrace{
\overbrace{m_1 < m_2}^{\mbox{atm}}
\ll
\overbrace{m_3 < m_4}^{\mbox{solar}}
}_{\mbox{LSND}}
\qquad \mbox{and} \qquad
\mbox{(B)}
\quad
\underbrace{
\overbrace{m_1 < m_2}^{\mbox{solar}}
\ll
\overbrace{m_3 < m_4}^{\mbox{atm}}
}_{\mbox{LSND}}
\;.
\end{equation}
In the case of these two mass spectra we have $r=2$ and thus
\begin{equation}\label{c2}
c_\alpha = \sum_{j=1,2} |U_{\alpha j}|^2 \quad (\alpha = e, \mu)\,.
\end{equation}
With the argument analogous to the one using 
$P^\odot_{\nu_e\to\nu_e}$ below Eq.(\ref{ineq}) one finds the
following constraint on the mixing matrix:
\begin{equation}\label{ABce}
\mbox{(A)} \quad c_e \leq a^0_e \,, \quad
\mbox{(B)} \quad c_e \geq 1-a^0_e \,.
\end{equation}


\begin{figure}[t]
\begin{center}
\mbox{
\subfigure[Scheme (A)]{\epsfig{file=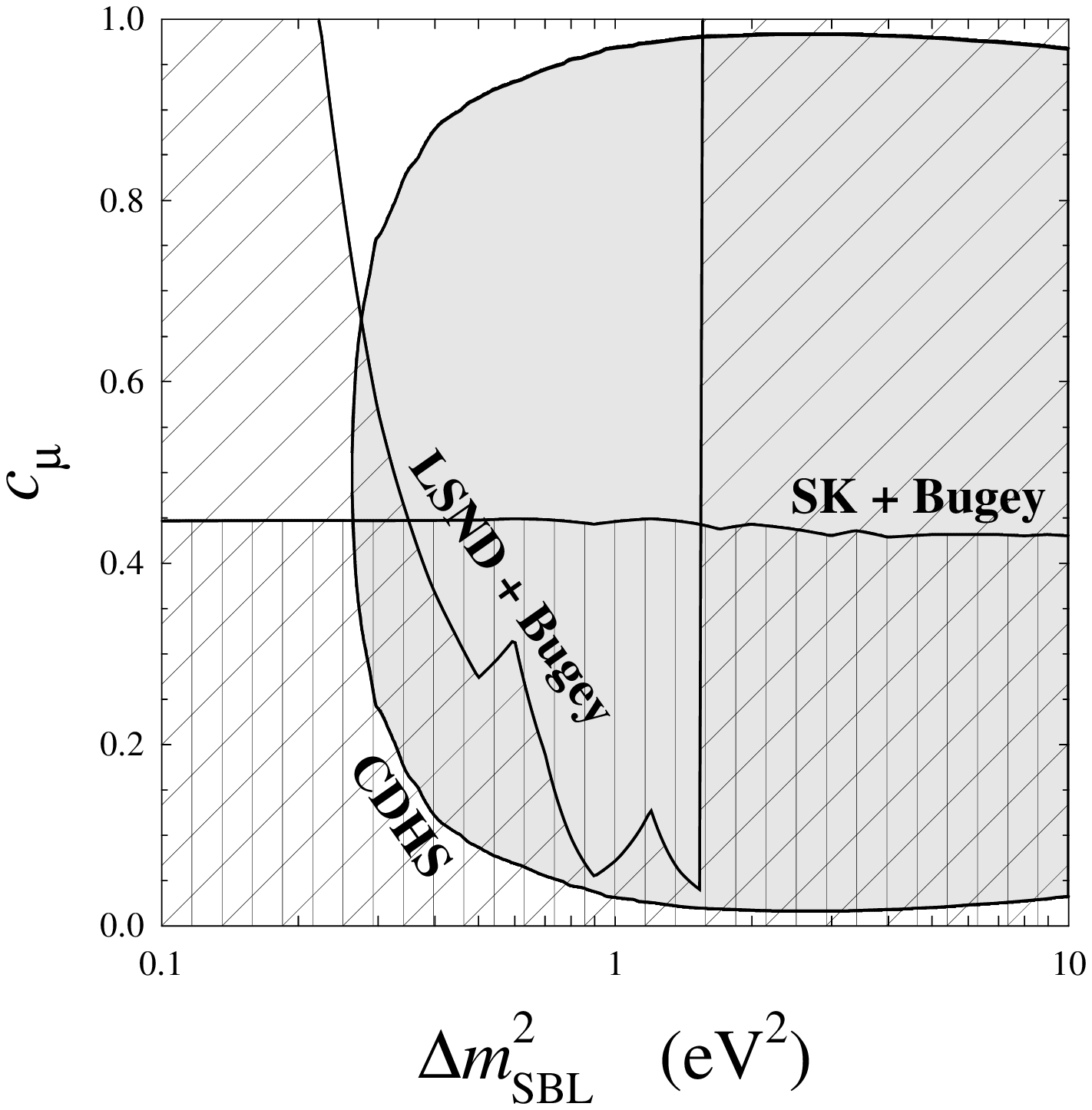,width=0.45\linewidth}}
\quad
\subfigure[Scheme (B)]{\epsfig{file=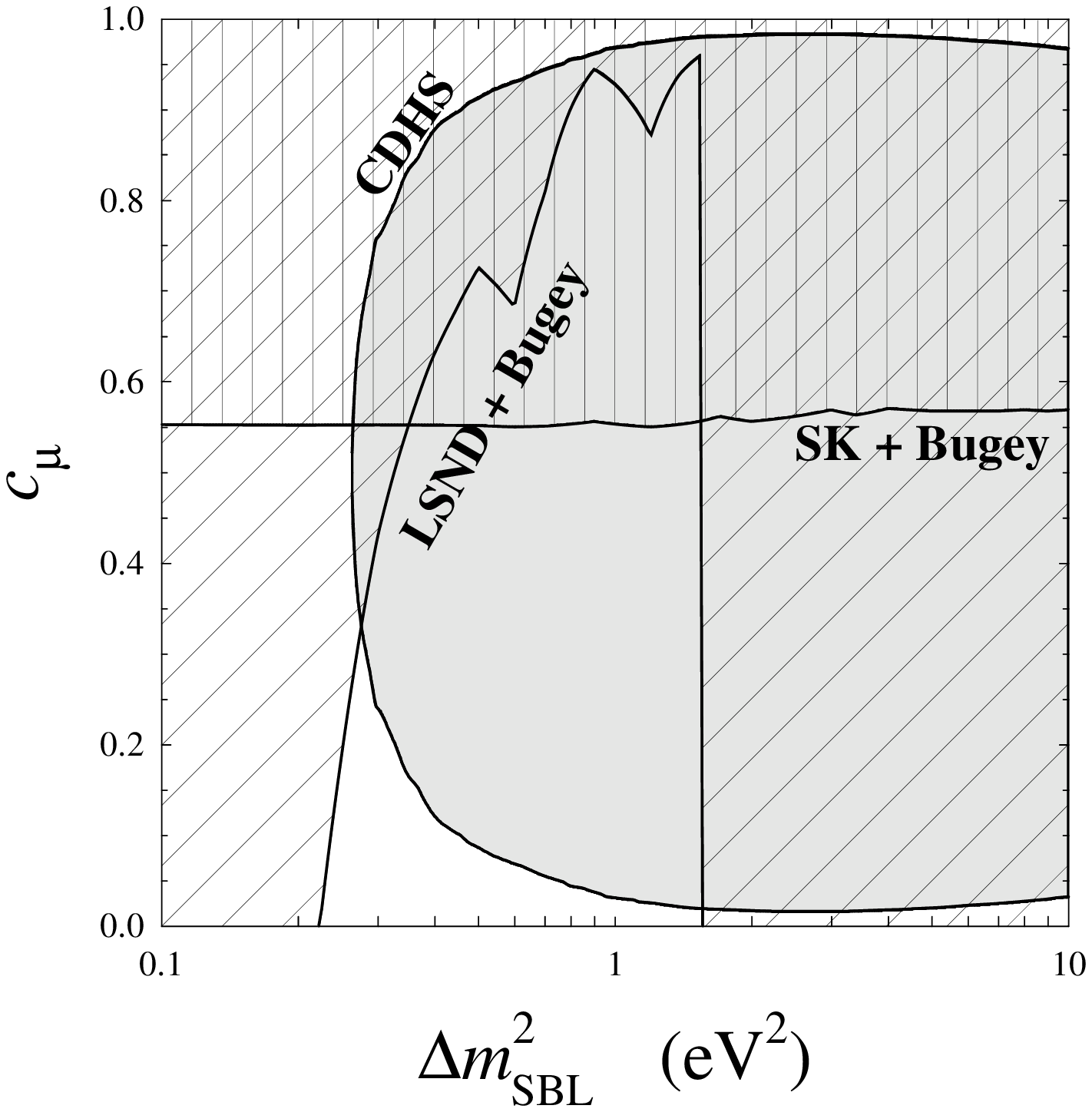,width=0.45\linewidth}}
}
\end{center}
\caption{The bounds for Schemes (A) and (B) analogous to the ones depicted in
Fig.~\ref{cmu} in the case of the hierarchical mass spectrum. The
white areas in Subfigs.~(a) and (b) show the allowed regions of
$c_\mu$ for Schemes (A) and (B), respectively. Subfig.~(b)
is obtained by reflection of all the curves in
Subfig.~(a) at the line $c_\mu = 1/2$.}
\label{cmuAB}
\end{figure}
We have to check that these mass spectra are compatible with the
results of all neutrino oscillation experiments. Going through the
same arguments as in the case of the hierarchical mass scheme in the
previous section, we have plotted the corresponding bounds a, b, c in
Fig.~\ref{cmuAB} for Scheme (A) (Subfig.~(a)) and (B) (Subfig.~(b)).
We observe that in these cases white areas (unshaded and
unhatched) are left \cite{BGGS99} which show the allowed ranges of 
$c_\mu$ in the $\Delta{m}^2_\mathrm{SBL}$--$c_\mu$ plane. Thus, 
Schemes (A) and (B) are compatible with all oscillation data.

\section{The sterile neutrino and big-bang nucleosynthesis}
\label{BBN}

Since we are discussing four neutrinos it is necessary to study the
compatibility of the preferred Schemes (A) and (B) with BBN. Whether
the effective number $N_\nu$ of light neutrinos relevant in BBN is
smaller than 4, is still debated in the literature.
An upper bound on $N_\nu$ depends, in particular, on the
primordial deuterium abundance $(D/H)_P$ for which conflicting
measurements exist. For the low value of $(D/H)_P$ the value of
$N_\nu$ should rather be close to 3 \cite{burles} 
whereas a high ratio $(D/H)_P$ allows also values of
$N_\nu$ around 4 \cite{olive}. In this section we inquire constraints
on the mixing matrix $U$ under the assumption of $N_\nu < 4$. In this case,
a large active -- sterile neutrino mixing seems to be excluded by
standard BBN with zero lepton number asymmetry 
(see Refs.~\cite{shi,OY97,BGGS98} and citations therein). 

In a simplified version, the amount of sterile neutrinos present at
BBN can be calculated using the differential equation \cite{kainulainen}
\begin{equation}\label{master}
\frac{dn_{\nu_s}}{dt}=
\frac{1}{2} \sum_{\alpha=e,\mu,\tau}
\langle P_{\nu_\alpha\to\nu_s} \rangle_{\mathrm{coll}}
\Gamma_{\nu_\alpha} (1-n_{\nu_s}) \,,
\end{equation}
where $n_{\nu_s}$ is the number density of the sterile neutrino relative
to the number density of an active neutrino in thermal equilibrium
($n_{\nu_s} \leq 1$) and the $\Gamma_{\nu_\alpha}$ are the total collision
rates of the active neutrinos \cite{enqvist}. The oscillation
probabilities in Eq.(\ref{master}) are averaged over the collision time
$t_\mathrm{coll} = 1/\Gamma_{\nu_e}$. Eq.(\ref{master}) is valid if
the oscillation time is smaller than the collision time and if in the time
evolution no resonance is encountered or a resonance is undergone
adiabatically. For a further discussion of
(\ref{master}) see Ref.~\cite{BGGS98}.

It turns out that in the time evolution in the early universe from a
temperature of around 100 MeV to a few MeV, when the active neutrinos 
decouple, in Scheme (A) there is no resonance, whereas in Scheme (B)
the time evolution goes through a non-adiabatic resonance. In the
latter case the Landau--Zener effect has to be used to estimate the
amount of sterile neutrinos produced at the resonance, instead of
using Eq.(\ref{master}). In this way the following constraint on $U$
can be derived \cite{BGGS98}:
\begin{equation}\label{cs}
N_\nu < 3.9 \; \Rightarrow \; \left\{
\begin{array}{rclc} 
  c_s & < 5 \times 10^{-3} & \mbox{(A)}\,, \\
1-c_s & < 5 \times 10^{-5} & \mbox{(B)}\,, 
\end{array} \right.
\end{equation}
where $c_s = \sum_{j=1,2} |U_{sj}|^2$.

Thus, concentrating on Scheme (A), from Eqs.(\ref{ABce}) and
(\ref{cs}) and from Fig.~\ref{cmuAB} we know which elements in the mixing 
matrix must be small in the rows pertaining to the neutrino flavours 
(types) $e$, $\mu$ and $s$. Consequently, also the small elements in
the $\tau$ row of $U$ are fixed. Symbolizing by $\circ$ small mixing
elements and by $\bullet$ large ones, we arrive at the following mixing
matrix: 
\begin{equation}\label{mixing-bbn}
\begin{array}{ccc}
&
\begin{array}{cccc} 1 & 2 & 3 & 4 \end{array}
& \\
\mbox{Scheme (A):} \quad U = \!\! &
\left( \begin{array}{cccc} 
\circ & \circ & \bullet & \bullet \\
\bullet & \bullet & \circ & \circ \\
\bullet & \bullet & \circ & \circ \\
\circ & \circ & \bullet & \bullet
\end{array} \right) & \!\!\!
\begin{array}{c} e \\ \mu \\ \tau \\ s \end{array}
\end{array}
\end{equation}
For Scheme (B) the analogous mixing matrix is obtained by the exchange
$1 \leftrightarrow 3$, $2 \leftrightarrow 4$ of the columns in
(\ref{mixing-bbn}). As a consequence, if $N_\nu < 4$,
the solar neutrino problem is solved $\nu_e \to \nu_s$ transitions in
Schemes (A) and (B), whereas the atmospheric neutrino anomaly by
$\nu_\mu\to\nu_\tau$ transitions \cite{OY97,BGGS98}.
Since a large mixing angle $\nu_e\to\nu_s$
transition as a solution of the solar neutrino puzzle is not
compatible with the solar neutrino data \cite{petcov}, this transition
must take place due to the small mixing angle MSW effect.

There is a debate in the literature if the constraint (\ref{cs}) can
be avoided by taking into account the effect of
a lepton number asymmetry in the
early universe. It rather seems that this is not possible with
the range (\ref{range}) of $\Delta m^2_\mathrm{SBL}$ determined by the LSND
experiment. For recent papers on this problem see Refs.~\cite{shi,Lasymm}.

\section{Predictions of the favoured Schemes (A) and (B)}

Schemes (A) and (B), either with or without the constraints from BBN,
allow to make predictions for LBL and SBL experiments, CP violation in
LBL experiments, $^3$H decay and $(\beta\beta)_{0\nu}$ decay. We will
not touch the subject of CP violation (see the papers in Ref. \cite{CP}).

\noindent
\textbf{LBL experiments:}\\
LBL neutrino oscillation experiments are sensitive to the so-called
``atmospheric $\Delta m^2$ range'' of $10^{-2}$--$10^{-3}$ eV$^2$. For
reactor experiments with $E \sim 1$ MeV this requires $L \sim 1$
km \cite{CHOOZ}, 
whereas in accelerator experiments with $E \sim 1$--10 GeV the
length $L$ of the baseline is of order of a few 100 to 1000 km 
\cite{KEK-SK,MINOS,ICARUS} (see Eq.(\ref{phase})).

Let us consider scheme (A) and neutrinos for definiteness. Then 
in vacuum the probabilities of
$ \nu_\alpha \to \nu_\beta $
transitions
in LBL experiments
are given by
\begin{equation}
P^{(\mathrm{LBL,A})}_{\nu_\alpha\to\nu_\beta}
=
\left|
U_{\beta1}
\,
U_{\alpha1}^{*}
+
U_{\beta2}
\,
U_{\alpha2}^{*}
\,
\exp \left( -i \frac{\Delta m^2_\mathrm{atm}L}{2E} \right)
\right|^2
+
\Bigg|
\sum_{k=3,4}
U_{{\beta}k}
\,
U_{{\alpha}k}^{*}
\Bigg|^2
\,.
\label{plab}
\end{equation}
This formula has been obtained from Eq.(\ref{prob})
by dropping terms with large phases being approximately
$\Delta m^2_\mathrm{SBL}L/2E$,
which do not contribute to the oscillation
probabilities averaged over the
neutrino energy spectrum.

\begin{figure}[t]
\begin{center}
\mbox{\epsfig{file=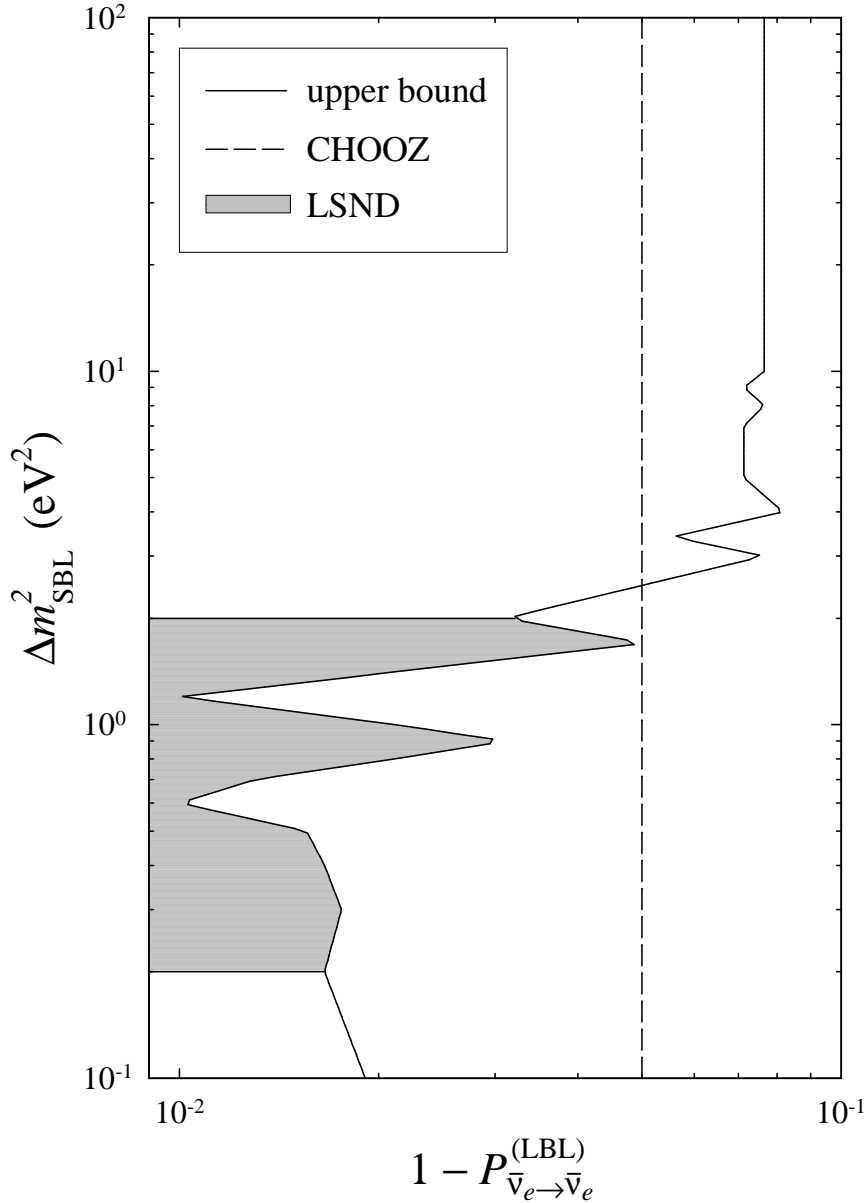,width=0.7\linewidth}}
\end{center}
\caption{ \label{peev} 
The upper bound (\ref{ee}) on 
$1-P^{(\mathrm{LBL})}_{\bar\nu_e\to\bar\nu_e}$. The shaded area
indicates the LSND range (\ref{range}) and the vertical dashed line
the experimental upper bound which has been 
obtained in the CHOOZ experiment.}
\end{figure}

From Eq.(\ref{plab}), with $\alpha = \beta$ and Eq.(\ref{c2}),
it follows immediately that
\begin{equation}
P^{(\mathrm{LBL,A})}_{\nu_\alpha\to\nu_\alpha} \geq (1-c_\alpha)^2 \,.
\end{equation}
Applying this inequality to LBL reactor experiments we obtain \cite{BGG97a}
\begin{equation}\label{ee}
1-P^{(\mathrm{LBL})}_{\bar\nu_e\to\bar\nu_e}
\leq a^0_e\, (2-a^0_e) \,.
\end{equation}
One can easily check that Eq.(\ref{ee}) holds for both Schemes (A) and
(B). In Fig.~\ref{peev} we have plotted this bound together with the
present experimental bound achieved in the CHOOZ experiment
\cite{CHOOZ}. The negative result of the CHOOZ experiment is in
agreement with the predictions of Schemes (A) and (B).

Considering now LBL transition probabilities and using the
Cauchy--Schwarz inequality for the first term on the right-hand side
of Eq.(\ref{plab}), we obtain
\begin{equation}
P^{(\mathrm{LBL,A})}_{\nu_\alpha\to\nu_\beta}
\leq
c_{\alpha}
\,
c_{\beta}
+
\frac{1}{4}
\,
A_{\alpha;\beta} \quad (\alpha \neq \beta)
\,.
\label{pab1}
\end{equation}
Whereas for the inequality (\ref{ee}) matter
corrections play no role due to the small energy of reactor neutrinos
and the distance $L \sim 1$ km of the detector from the source,
such corrections have to be taken into account to derive a realistic bound
from Eq.(\ref{pab1}) in order to apply it to LBL accelerator
experiments \cite{KEK-SK,MINOS,ICARUS}. For a derivation of a
matter-corrected, scheme-independent upper bound from Eq.(\ref{pab1})
see Ref. \cite{BGG97a}. For the MINOS and ICARUS
experiments this upper bound on the $\nu_\mu\to\nu_e$ transition
probability decreases from around 0.1 to 0.03 when $\Delta
m^2_\mathrm{SBL}$ varies from 0.2 to 2 eV$^2$ (\ref{range})
\cite{BGG97a}. However, the sensitivity of these experiments is much
better than this bound. For
the KEK to Super-Kamiokande LBL experiment the upper bound on the same
transition is rather 0.04 at most \cite{BGG97a}. A similar stringent
bound can be derived on the probability of $\nu_e\to\nu_\tau$
transitions.  

\noindent
\textbf{SBL $\nu_\mu\to\nu_\tau$ transitions:}\\
If $N_\nu < 4$ holds (see previous section), then the quantity $c_s$
is very small (see Eq.(\ref{cs})). In this case it can be shown that
\begin{equation}
\left. \begin{array}{cr} \mbox{(A)} &   c_s \to 0 \\
                         \mbox{(B)} & 1-c_s \to 0 \end{array} \right\}
\; \Rightarrow \; A_{\mu;\tau} \leq (a^0_e)^2
\end{equation}
is valid \cite{BGGS98}. Due to the smallness of $a^0_e$ (see
Fig.~\ref{a0plot}) the transition amplitude $A_{\mu;\tau}$ is below
$10^{-3}$ \cite{BGGS98}, which also serves as a check for the
validity of the BBN constraint on $U$.

\noindent
\textbf{$^3$H decay:}\\
Let us assume that $m_{1,2} \ll m_{3,4}$. Then one easily derives the
relations \cite{BGG96}
\begin{equation}
\begin{array}{cl} \mbox{(A)} & m_\nu(^3\mbox{H}) \simeq m_3 \simeq m_4
\,, \\            \mbox{(B)} & m_\nu(^3\mbox{H}) \lesssim a^0_e m_4 
\end{array}
\end{equation}
for the mass $m_\nu(^3\mbox{H})$ measured in tritium decay.
Since in Scheme (A) one has 
$m_{3,4} \simeq (\Delta m^2_\mathrm{SBL})^{1/2}$, it might be possible
in the future to see a neutrino mass in tritium decay, whereas this
mass effect is suppressed in Scheme (B).

\noindent
\textbf{$(\beta\beta)_{0\nu}$ decay:}\\
If neutrinos are of Majorana nature, neutrinoless double-beta decay 
proceeds via the effective Majorana neutrino mass (see
Ref.~\cite{mohapatra} for other mechanisms)
\begin{equation}\label{m}
|\langle m \rangle |= \left|\sum_{j=1}^4 U^2_{ej} m_j \right| \,,
\end{equation}
which can thus be related to present experimental information. 
Making the same assumptions about the neutrino masses as in the
previous paragraph, it is easy to show that in Scheme (A) the relation
\begin{equation}\label{mbound}
\sqrt{1 - \sin^2 2\theta_\mathrm{solar}} \sqrt{\Delta m^2_\mathrm{SBL}} 
\lesssim | \langle m \rangle | \lesssim  \sqrt{\Delta m^2_\mathrm{SBL}} 
\end{equation}
holds \cite{betabeta}, where $\theta_\mathrm{solar}$ is the mixing
angle relevant for solar neutrino oscillations. Note that from the
range (\ref{range}) it follows that
\begin{equation}\label{sqrtrange}
0.5 \; \mbox{eV} \lesssim
\sqrt{\Delta m^2_\mathrm{SBL}}
\lesssim 1.4 \; \mbox{eV}. 
\end{equation}
Thus, at least the upper bound in Eq.(\ref{mbound}) is in the reach of 
present experiments \cite{faessler,baudis}.
At present a very stringent bound exists from the $^{76}$Ge experiment
\cite{baudis} with $| \langle m \rangle | \lesssim 0.2 \div 0.6$ eV (see
also the references cited in Ref.~\cite{baudis}). Note that for the
small mixing angle MSW solution of the solar neutrino puzzle, which is
favoured by BBN, one has
\begin{equation}
| \langle m \rangle | \simeq  \sqrt{\Delta m^2_\mathrm{SBL}} \,.
\end{equation}

\section{Conclusions}

In this report we have discussed the possible form of the
neutrino mass spectrum
that can be inferred from
the results of all
neutrino oscillation experiments,
including solar and atmospheric neutrino
experiments. The crucial input are the three indications in favour of
neutrino oscillations given by the solar neutrino data, the
atmospheric neutrino anomaly and the result of the LSND
experiment, and also the negative results of the SBL disappearance
experiments. These indications, which all pertain to different scales
of neutrino mass-squared differences, require that apart from the 
three well-know neutrino flavours at least one additional sterile
neutrino (without couplings to the $W$ and $Z$ bosons) must exist.
In our investigation we have
assumed that there is one sterile neutrino and that the
4-neutrino mixing matrix (\ref{mixing}) is unitary.
We have considered all possible
schemes with four massive neutrinos
which provide 
three scales of $\Delta{m}^2$ (see Fig.~\ref{4spectra}).

The main points of our discussion can be summarized as follows:
\begin{itemize}
\item
The data prefer the non-hierarchical mass spectra (A) and (B) (see
Fig.~\ref{4spectra} and Eq.(\ref{AB}))
with two pairs of close masses separated by a mass difference
of the order of 1 eV necessary for a description of the LSND
result. In Scheme (A), the quantity $|U_{e1}|^2+ |U_{e2}|^2$ is small
and $|U_{\mu 1}|^2+ |U_{\mu 2}|^2$ is close to 1, and vice versa in
Scheme (B).
\item
The solar neutrino problem is preferably solved by 
$\nu_e \to \nu_\tau, \nu_s$ and the atmospheric neutrino
anomaly by $\nu_\mu \to \nu_\tau, \nu_s$ transitions. If the effective
number $N_\nu$ of neutrinos relevant in BBN is smaller
than 4, then standard BBN leads to 
small mixing angle MSW $\nu_e \to \nu_s$
transitions as the solution of the solar neutrino problem and to
$\nu_\mu \to \nu_\tau$ transitions in atmospheric neutrinos.
\item
Again with $N_\nu < 4$, $\nu_\mu \to \nu_\tau$ transitions are
strongly suppressed in SBL neutrino oscillations. Note that 
in the case of $N_\nu < 4$ 
all SBL neutrino oscillations are small or suppressed.
\item
In LBL neutrino oscillations, it follows from Schemes (A) and (B) that
the transitions $\nu_e \to \nu_\alpha \; (\alpha \neq e)$ and 
$\nu_\mu \to \nu_e$ are suppressed.
\item
Schemes (A) and (B) could in principle be distinguished in $^3$H and
$(\beta\beta)_{0\nu}$ decays, because in Scheme (A) 
neutrino mass effects are
expected, whereas in Scheme (B) such effects 
are suppressed. Note that in Scheme (A) with the small mixing
angle MSW solution of the solar neutrino problem, which is preferred
by standard BBN (see Section \ref{BBN}), one gets 
$| \langle m \rangle | \simeq \sqrt{\Delta m^2_\mathrm{SBL}} \gtrsim
0.5$ eV for the effective Majorana mass relevant in
$(\beta\beta)_{0\nu}$ decay (see Eqs.(\ref{mbound}) and
(\ref{sqrtrange})). Such a large value for $| \langle m
\rangle |$ should be close to discovery.
\end{itemize}
 
Finally, we want to remark that the most crucial input in our
discussion is the result of the LSND experiment. This result will be checked
by the approved MiniBooNE experiment, which will begin data taking in
2001 \cite{conrad,BooNE}. The SNO experiment, which is expected to
announce the first results in 2000, will test the hypothesis of
oscillations of solar neutrinos into sterile neutrinos \cite{SNO}.
It could thus deliver a very important further piece of evidence in
favour of the sterile neutrino and thus indirectly also
check the BBN constraint on the 4-neutrino mixing matrix.

\section*{Acknowledgements}
The author would like to thank the organizers of the school for their
hospitality and the stimulating and pleasant
atmosphere. Furthermore, he is very grateful to C. Giunti for
updating or preparing some of the figures presented in this report.

\end{document}